\documentclass[aps,prb,twocolumn,superscriptaddress]{revtex4-2}
\usepackage[dvipdfmx]{graphicx}
\usepackage{color}
\usepackage{ulem}
\begin{document}


\title{Three-dimensional electronic structure in ferromagnetic $\textrm{Fe}_3\textrm{Sn}_2$ with breathing kagome bilayers}


\author{Hiroaki Tanaka}
\affiliation{ISSP, University of Tokyo, Kashiwa, Chiba 277-8581, Japan}

\author{Yuita Fujisawa}
\affiliation{Okinawa Institute of Science and Technology Graduate University, Okinawa 904-0495, Japan}

\author{Kenta Kuroda}
\email{kuroken224@issp.u-tokyo.ac.jp}
\affiliation{ISSP, University of Tokyo, Kashiwa, Chiba 277-8581, Japan}

\author{Ryo Noguchi}
\affiliation{ISSP, University of Tokyo, Kashiwa, Chiba 277-8581, Japan}

\author{Shunsuke Sakuragi}
\affiliation{ISSP, University of Tokyo, Kashiwa, Chiba 277-8581, Japan}

\author{C\'edric Bareille}
\affiliation{ISSP, University of Tokyo, Kashiwa, Chiba 277-8581, Japan}

\author{Barnaby Smith}
\affiliation{Okinawa Institute of Science and Technology Graduate University, Okinawa 904-0495, Japan}

\author{Cephise Cacho}
\affiliation{Diamond Light Source, Harwell Campus, Didcot OX11 0DE, United Kingdom}

\author{Sung Won Jung}
\affiliation{Diamond Light Source, Harwell Campus, Didcot OX11 0DE, United Kingdom}

\author{Takayuki Muro}
\affiliation{Japan Synchrotron Radiation Research Institute (JASRI), 1-1-1 Kouto, Sayo, Hyogo 679-5198, Japan}

\author{Yoshinori Okada}
\affiliation{Okinawa Institute of Science and Technology Graduate University, Okinawa 904-0495, Japan}

\author{Takeshi Kondo}
\email{kondo1215@issp.u-tokyo.ac.jp}
\affiliation{ISSP, University of Tokyo, Kashiwa, Chiba 277-8581, Japan}
\affiliation{Trans-scale Quantum Science Institute, University of Tokyo, Bunkyo-ku, Tokyo 113-0033, Japan}


\date{\today}

\begin{abstract}
A large anomalous Hall effect (AHE) has been observed in ferromagnetic $\textrm{Fe}_3\textrm{Sn}_2$ with breathing kagome bilayers.
To understand the underlying mechanism for this, we investigate the electronic structure of $\textrm{Fe}_3\textrm{Sn}_2$ by angle-resolved photoemission spectroscopy (ARPES).
In particular, we use both vacuum ultraviolet light (VUV) and soft x ray (SX), which allow surface-sensitive and relatively bulk-sensitive measurements, respectively, and distinguish bulk states from surface states, which should be unlikely related to the AHE.
While VUV-ARPES observes two-dimensional bands mostly due to surface states, SX-ARPES reveals three-dimensional band dispersions with a periodicity of the rhombohedral unit cell in the bulk.
Our data show a good consistency with a theoretical calculation based on density functional theory, suggesting a possibility that $\textrm{Fe}_3\textrm{Sn}_2$ is a magnetic Weyl semimetal.
\end{abstract}


\maketitle




The translational symmetry of a crystal allows its eigenstates to be labeled by Bloch wave vectors in a reciprocal space \cite{Ashcroft}, and the Berry curvature \cite{Berry1984} defined there could cause several anomalous physical behaviors.
The integration of the Berry curvature over the occupied states leads to an intrinsic anomalous Hall effect (AHE), which becomes nonzero under broken time-reversal symmetry.
Moreover, in a two-dimensional system, the integration over the Brillouin zone (BZ) for each different band is quantized to an integer \cite{PhysRevLett.49.405}, which is called the Chern number and has attracted growing interests as a topological invariant \cite{RevModPhys.82.3045}. 
Non-zero Chern number is realized, for example, in the honeycomb lattice model with gapped Dirac fermionic states \cite{PhysRevLett.95.226801}. 
An extension of these two-dimensional states to a three-dimensional crystal, under the breaking of time-reversal or space-inversion symmetry, which resolves spin degeneracy, is a  Weyl semimetal \cite{PhysRevLett.107.186806}, which has been intensively studied theoretically \cite{Huang2015,Soluyanov2015, PhysRevLett.117.066402, Yang_2017} and experimentally \cite{Xu2015, PhysRevX.5.031013, Lv2015, Kuroda2017, PhysRevLett.122.176402, Liu1282}.

\begin{figure}
\includegraphics[width=88mm]{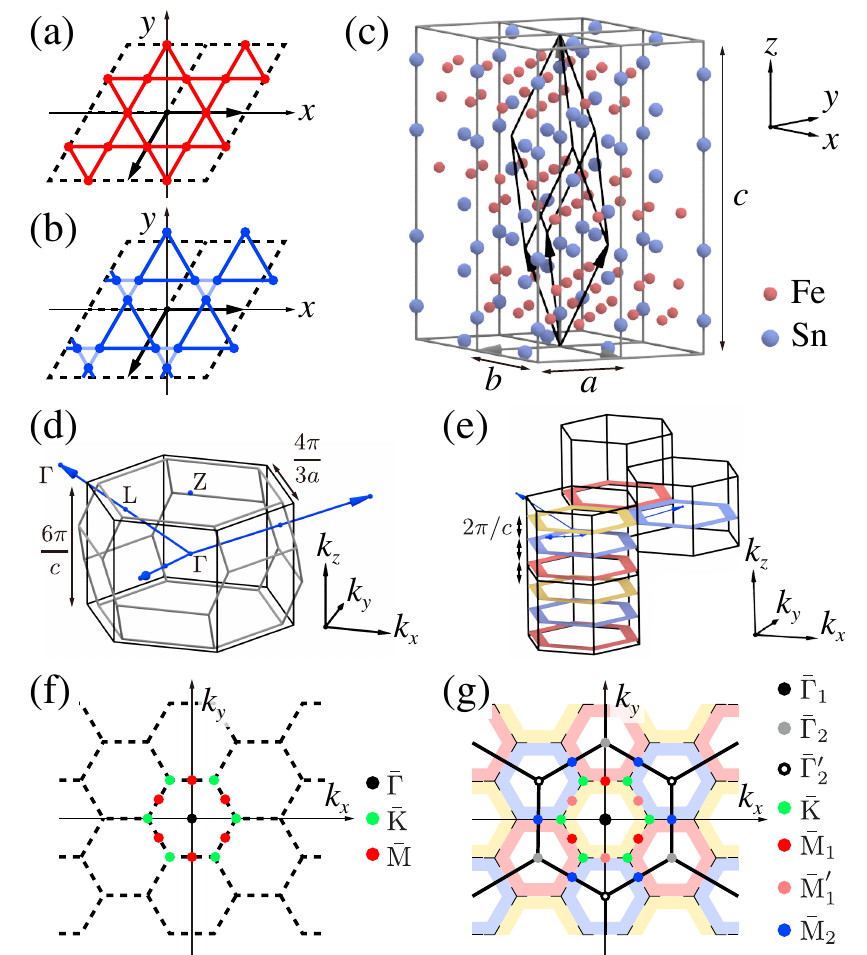}%
\caption{\label{Figure: Structure} Structure of kagome lattice and $\textrm{Fe}_3\textrm{Sn}_2$. (a), (b) Ideal and breathing kagome lattice, respectively. (c) Crystal structure of $\textrm{Fe}_3\textrm{Sn}_2$ with conventional unit cells (gray lines) and primitive unit cell (black lines). The lattice constants of the conventional unit cell were determined to be $a=b=5.35\ \textrm{\AA}$ and $c=19.82\ \textrm{\AA}$ by x-ray diffraction measurements. (d) Brillouin zone (gray lines), hexagonal column unit cell (black lines), and unit vectors (blue arrows). (e) Stacking of the hexagonal column unit cells. (f), (g) Periodicity for the two-dimensional structure (black dashed lines) and that for three-dimensional one (black solid lines), respectively. High-symmetry points are marked by circles.}
\end{figure}

Graphene with Dirac fermionic states yielded by the honeycomb lattice \cite{PhysRev.109.272} is known as a model case in which topological phenomena are well described. As another possible system hosting Dirac fermionic states, the two-dimensional ideal kagome lattice [Fig.\ \ref{Figure: Structure}(a)] has been proposed \cite{PhysRevB.80.113102}.
From this perspective, ferromagnetic $\textrm{Fe}_3\textrm{Sn}_2$ with the kagome lattice has been of recent interest in condensed matter physics.
This compound is actually not composed of the ideal kagome lattice, but of the breathing kagome lattice [Fig.\ \ref{Figure: Structure}(b)], which is formed by upward and downward triangles with different sizes and therefore has two atomic bond lengths.
Despite it, $\textrm{Fe}_3\textrm{Sn}_2$ exhibits various attractive properties, such as a large AHE \cite{PhysRevB.94.075135,Ye2018}, a topological Hall effect \cite{Fe3Sn2THE}, large tunability of magnetization \cite{Yin2018}, gapped (massive) Dirac states \cite{Ye2018}, and flat bands near the Fermi energy \cite{PhysRevLett.121.096401}.
$\textrm{Fe}_3\textrm{Sn}_2$ has been, therefore, expected to provide a new platform to study topological physics in the kagome network with broken time-reversal symmetry.

Previously, the origin of the AHE in $\textrm{Fe}_3\textrm{Sn}_2$ was proposed to be two-dimensional Dirac fermionic states with band gaps opened by spin-orbit coupling (SOC).
We note here that, while only SOC is assumed to contribute to the band gaps in the previous study, our analysis demonstrates that the breathing effect can also open similar gaps \cite{PhysRevB.99.165141}.
Very importantly, the resulting Berry curvature should have different characteristics depending on the origins of the gaps: SOC or the breathing effect (Supplemental Material Notes 1 and 2 \cite{Supplemental}).
Based on group theory \cite{PhysRev.50.58,PhysRev.96.280, OPECHOWSKI1940552}, our analysis without SOC shows two degeneracies in a bilayer at the $K$ points, as claimed in Ref.\ \cite{Ye2018}.
Nevertheless, this result does not necessarily mean that the breathing effect [Supplemental Material Fig.\ S4(b) \cite{Supplemental}] disappears. 

We also point out that angle-resolved photoemission spectroscopy (ARPES) experiments with vacuum ultraviolet light (VUV), previously used \cite{Ye2018}, could become so surface sensitive as to capture only surface states because of the short mean free path of the photoelectrons excited \cite{POWELL19991}.
The shortcoming of VUV-ARPES could be, however, solved by using soft x ray (SX) as a photon source; the mean free path becomes longer than that of VUV-ARPES, making SX-ARPES more suited to study bulk properties of a crystal \cite{STROCOV20191}.
In this Rapid Communication, we utilize both surface-sensitive VUV-ARPES and relatively bulk-sensitive SX-ARPES to distinguish the bulk and surface electronic structures of $\textrm{Fe}_3\textrm{Sn}_2$.
In contrast to VUV-ARPES, our SX-ARPES experiments reveal three-dimensional band dispersions, moderately consistent with those of Kohn-Sham orbitals in density functional theory (DFT) \cite{RevModPhys.71.1253}.
This coincidence suggests that the presence of Weyl semimetallic states could be the origin of the AHE.

$\textrm{Fe}_3\textrm{Sn}_2$ crystals were grown by the chemical vapor transport method.
VUV-ARPES measurements were performed at I05 ARPES beamline of Diamond Light Source \cite{Diamond}.
The photon energies used ranged from 50 to 120 eV, and the overall energy resolution was less than 20 meV.
SX-ARPES measurements were performed at BL25SU of SPring-8 \cite{SPring-8}.
The photon energies used ranged from 380 to 620 eV, and the overall energy resolution was less than 60 meV.
In both experiments, the samples were cleaved \textit{in} \textit{situ} along the (001) plane at an ultra-high vacuum of $\sim 3\times10^{-8}\ \textrm{Pa}$, and the temperature for measurements was set at about 60 K.
In our DFT calculations without SOC, we used {\sc quantum espresso} code \cite{Giannozzi_2009, Giannozzi_2017} and ultrasoft pseudopotentials \cite{PhysRevB.41.1227, PSlibrary}.
Atomic positions were optimized using a $16\times16\times16$ $k$ mesh, and then self-consistent calculations were performed with a $32\times32\times32$ $k$ mesh.
During the calculations, the size of the primitive unit cell was fixed to that experimentally determined.

The reciprocal space associated with the crystal structure needs to be carefully taken into account to understand the ARPES signature.
$\textrm{Fe}_3\textrm{Sn}_2$ is composed of a Fe-Sn kagome bilayer and a Sn honeycomb lattice stacked along the $z$ direction [Fig.\ \ref{Figure: Structure}(c)].
While the conventional unit cell is shaped by the quadrangular prism [gray lines in Fig.\ \ref{Figure: Structure}(c)], a smaller rhombohedral primitive unit cell is also defined (black lines).
Here we consider the periodicity in the reciprocal space for the primitive unit cell, which should be observed for the bulk bands by ARPES. 
For ease of understanding, we use a hexagonal column [black lines in Fig.\ \ref{Figure: Structure}(d)] as a unit cell of the reciprocal space, instead of the BZ (gray lines).
The hexagonal columns are simply stacked along the $k_z$ direction, whereas the ones touched on the side surfaces are shifted by $2\pi/c$ from each other [Fig.\ \ref{Figure: Structure}(e)].
The colored hexagonal frames in Fig.\ \ref{Figure: Structure}(e) can be neglected in a two-dimensional electronic structure (or surface states), which thus expects a small hexagonal periodicity, as presented in Fig.\ \ref{Figure: Structure}(f).
In contrast, the colored frame should be taken into account for a three-dimensional electronic structure (or bulk states), leading to a longer periodicity [Fig.\ \ref{Figure: Structure}(g)]. 
The rhombohedral unit cell, therefore, enables us to distinguish between two- and three-dimensional electronic structures not only from $k_z$ dispersion but also from in-plane periodicity.

Here we provide a theoretical consideration of the electronic states allowed by the crystal property of $\textrm{Fe}_3\textrm{Sn}_2$ (see Notes 1--3 \cite{Supplemental} for details).
Assuming two-dimensionality, the nearest-neighbor tight-binding model for the breathing kagome lattice presents cone-shaped dispersions centered at the $\bar{K}$ points, which become gapped even without SOC [Ref.\ \cite{PhysRevB.99.165141} and Fig.\ S1(c) \cite{Supplemental}].
The band gap opened either by the breathing effect or SOC [Refs.\ \cite{PhysRevB.99.165141,Ye2018}, and Fig.\ S1(b) \cite{Supplemental}] can generally be understood by group theory [Note 1 \cite{Supplemental}].
Significantly, the Berry curvature generated by the breathing effect [Fig.\ S2(b) \cite{Supplemental}] is expressed as an odd function, whereas that by SOC [Fig.\ S2(c) \cite{Supplemental}] as an even function.
Hence, the values of the Berry curvature are not uniquely determined by the size of the band gap.

This argument for the two-dimensional case can be extended to that for the three-dimensional case, in which the group theoretical analysis similarly confirms that electronic states have two degeneracies in a breathing bilayer at the $K$ points without SOC (Note 3 \cite{Supplemental}).
Importantly, this result, however, does not necessarily mean that massless Dirac fermionic states must be generated [Fig.\ S4(b) \cite{Supplemental}].
Moreover, these arguments based on the conventional unit cell are not straightforward for the understanding of the physical properties of $\textrm{Fe}_3\textrm{Sn}_2$ such as ARPES spectra, because they have a periodicity corresponding to the primitive unit cell.
Hereafter, we investigate the electronic structure observed by ARPES and discuss it based not on the conventional unit cell associated with the kagome lattice but on the rhombohedral primitive unit cell, which is the essential structure.

\begin{figure}
\includegraphics[width=80mm]{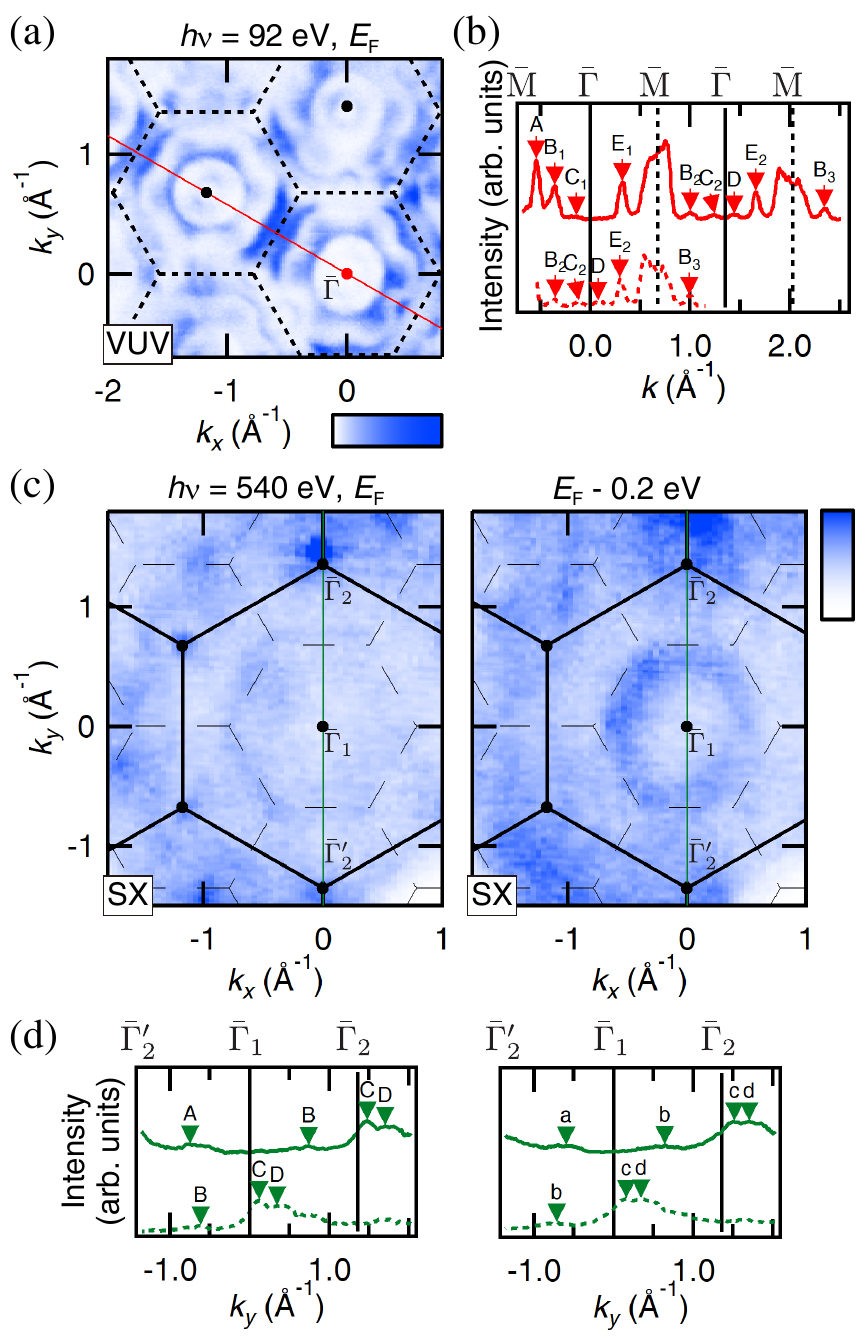}%
\caption{Constant-energy mappings of $\textrm{Fe}_3\textrm{Sn}_2$. (a) Fermi surface mapping measured by VUV-ARPES. The dashed lines represent the periodicity of the two-dimensional structure. (b) Momentum distribution curve (MDC) along the red line on (a), and that shifted by one period along the momentum direction. (c) Constant-energy mappings at $E_\mathrm{F}$ (left panel) and $E_\mathrm{F}-0.2\ \textrm{eV}$ (right panel) measured by SX-ARPES. The dashed and solid lines represent the periodicities of the two- and three-dimensional structures, respectively. (d) MDCs along the green lines on (c) and those shifted by one period of the two-dimensional periodicity [dashed lines in (c)].}
\label{Figure: Constant-energy} 
\end{figure}

\begin{figure}
\includegraphics[width=80mm]{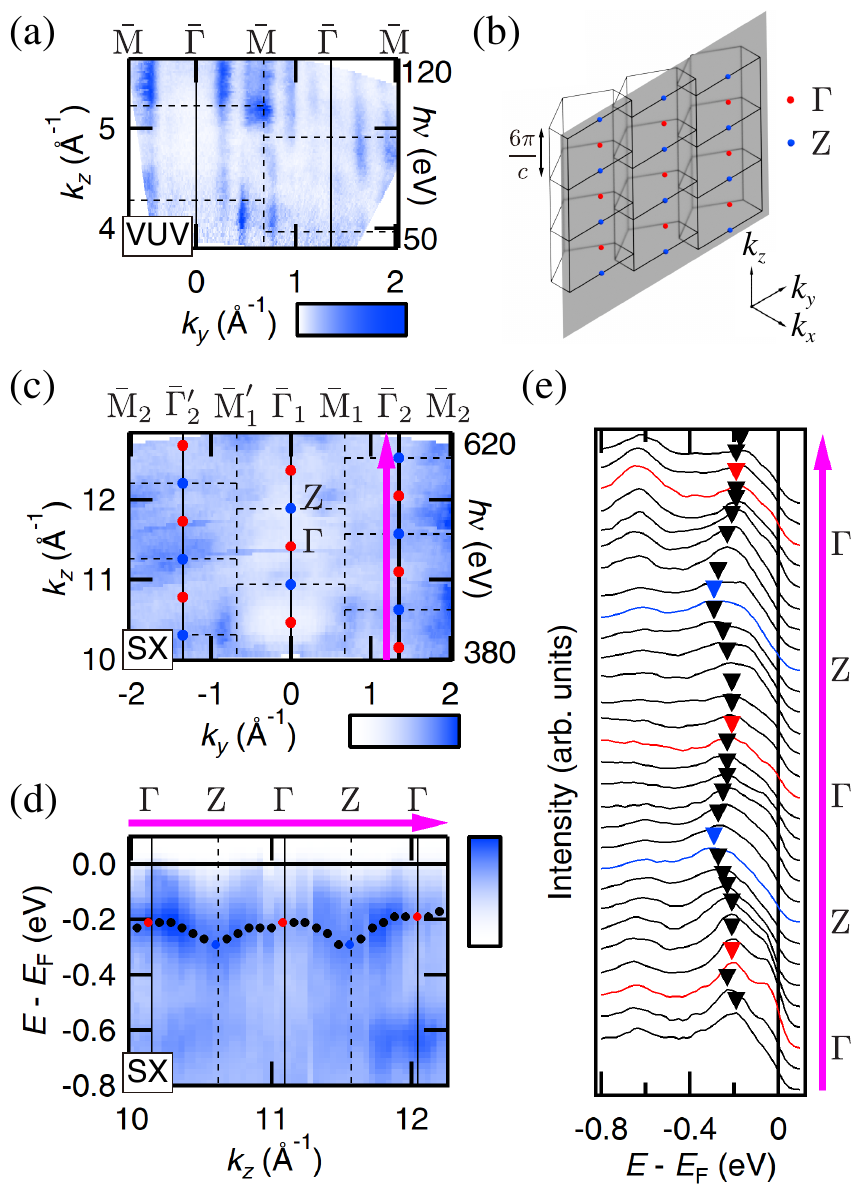}%
\caption{\label{Figure: kz} $k_z$ dispersion of electronic states in $\textrm{Fe}_3\textrm{Sn}_2$. (a) $k_y$-$k_z$ mapping at the Fermi energy measured by VUV-ARPES. (b) Cross-section at $k_x=0$ of the stacked hexagonal unit cells. (c) The $k_y$-$k_z$ mapping at $E_\mathrm{F}-0.4\ \textrm{eV}$ measured by SX-ARPES. (d), (e) Energy-momentum mapping along the $\bar{\Gamma}_2$ line [the pink arrow in (c)] and the corresponding energy distribution curves, respectively. For clarity, the spectra at the $\Gamma$ and $Z$ points are colored by red and blue, respectively. The spectral peaks are marked by triangles, and the corresponding energy states are plotted by circles in (d).}
\end{figure}

Figure \ref{Figure: Constant-energy}(a) plots the Fermi surface map along $k_x$-$k_y$ obtained by VUV-ARPES.  The periodicity of ARPES intensities is found to be short, agreeing with that for surface states; while some traces of bulk states are partially obtained at particular binding energies (Note 5 \cite{Supplemental}), these are not significant.
The feature of surface states is more clearly identified in Fig.\ \ref{Figure: Constant-energy}(b), which extracts the momentum distribution curve (MDC) across the $\bar{\Gamma}$ point [a red line in Fig.\ \ref{Figure: Constant-energy}(a)].
Along the momentum cut, spectral peaks corresponding to $k_\mathrm{F}$ positions periodically appear, as confirmed by a good matching between the original MDC and that shifted by one period (the distance between adjacent $\bar{\Gamma}$ points).
In sharp contrast, the Fermi surface map by SX-ARPES [the left panel of Fig.\ \ref{Figure: Constant-energy}(c)] exhibits a much longer periodicity: The large hexagonal pocket centered at the $\bar{\Gamma}_1$ point is different from the much smaller circular pockets at the $\bar{\Gamma}_2$ points.
These features become much clearer at slightly higher binding energy ($E_\mathrm{F}-0.2\ \textrm{eV}$), as demonstrated in the right panel of Fig. 2(c). The MDCs across $\bar{\Gamma}$ [green lines in Fig.\ \ref{Figure: Constant-energy}(c)] also confirm that SX-ARPES intensities lack signatures with the two-dimensional periodicity [Fig.\ \ref{Figure: Constant-energy}(d)], rather being compatible with bulk states with a longer periodicity.

\begin{figure*}
\includegraphics[width=160mm]{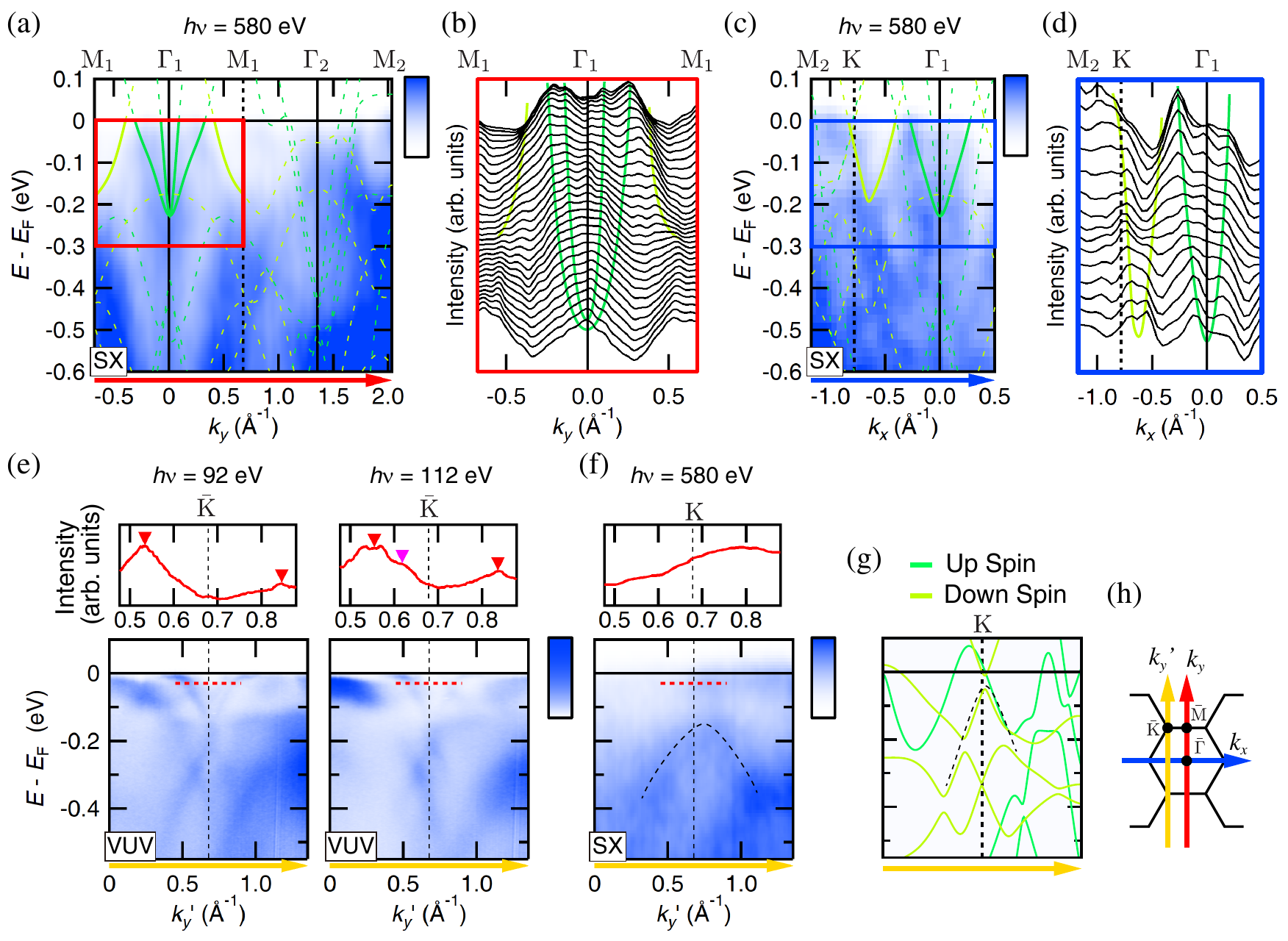}%
\caption{\label{Figure: Ek, DFT} Comparison between ARPES data and the dispersions of DFT calculations. (a) Energy-momentum mapping measured along the $k_y$ direction [a red arrow in (h)]. The dispersions of the DFT calculations (green and yellow green curves) are overlaid and those agreeing well with the ARPES results are particularly emphasized with thick curves. (b) Momentum distribution curves (MDCs) extracted from the red rectangular region in (a). Overlaid curves are guides for the eye, which correspond to the emphasized DFT dispersions. (c) Energy-momentum mapping  along the $k_x$ direction [a blue arrow in (h)], with overlaid DFT dispersions. (d) MDCs extracted from the blue rectangular region in (c), with schematic band dispersions plotted as guides for the eye. (e) Energy-momentum maps (bottom panels) obtained by VUV-ARPES along the yellow arrow in (h), exhibiting massive Dirac cones centered at the $\bar{K}$ point. 
In the top panels, MDCs at $E_\mathrm{F}-0.03\ \textrm{eV}$ (red dashed lines in the mappings) are extracted. The spectral peaks for outer and inner cones are marked by red and pink triangles, respectively. (f) SX-ARPES results of energy-momentum mapping (bottom panel) and MDC (top panel) along the same momentum cut as in (e).
(g) DFT band dispersions in the same momentum cut as in (e) and (f). Cone-shaped dispersions are traced with dashed curves in (f) and (g). (h) Schematic representing the two-dimensional periodicity with momentum cuts(arrows) along which the above ARPES mappings are taken.}
\end{figure*}

The periodicity of the band structure is further investigated along the $k_z$ direction by changing the photon energy ($h\nu$) in Fig.\ \ref{Figure: kz}.
For the data measured by VUV-ARPES, the $k_z$ dispersion is almost negligible [Fig.\ \ref{Figure: kz}(a)]; while the spectral intensities modulate with photon energy due to the matrix element effects, the peak positions of MDCs are unchanged [Fig.\ S7(b) \cite{Supplemental}], indicating that these spectra have dominant contribution from the surface states.
In contrast, the results of SX-ARPES display a clear $k_z$ dispersion [Figs.\ \ref{Figure: kz}(c)--(e)] with the periodicity for the bulk state, which is illustrated by misaligned rectangles along $k_y$-$k_z$ [Fig.\ \ref{Figure: kz}(b)].
In Fig.\ \ref{Figure: kz}(d), we extract the energy-momentum map along $k_z$ across the $\Gamma$ and Z points, and can indeed confirm the periodic oscillation expected for the bulk states by tracking a spectral peak position (filled circles) for each energy distribution curve [Fig.\ \ref{Figure: kz}(e)].

To further understand the bulk band structure in $\textrm{Fe}_3\textrm{Sn}_2$, here we compare the ARPES results by SX-ARPES with the dispersions of Kohn-Sham orbitals in DFT along the momentum cuts represented in Fig.\ \ref{Figure: Ek, DFT}(h).
The $\Gamma$ point in the three-dimensional BZ can be accessed at $h\nu=580\ \mathrm{eV}$ in the SX-ARPES measurements.
Figures \ref{Figure: Ek, DFT}(a) and \ref{Figure: Ek, DFT}(c) show the energy-momentum maps along the $k_y$ and $k_x$ directions, respectively.
The curvature plot \cite{doi:10.1063/1.3585113} of a wide-energy map clearly shows band dispersions, which are complex as seen in the DFT calculations [Fig.\ S8(b) \cite{Supplemental}]. Here we focus on several bands near $E_\mathrm{F}$ around the $\Gamma_1$ point and obtained consistency with the DFT band dispersions, which are emphasized by thick curves. This agreement is more clearly demonstrated in Figs.\ \ref{Figure: Ek, DFT}(b) and \ref{Figure: Ek, DFT}(d) by extracting MDCs, in which one can trace two electron bands and one hole band centered at the $\Gamma_1$ point (see overlaid guides).
Although the resolution of SX-ARPES is not perfect, this result reasonably supports a theoretical prediction by \textit{ab initio} calculations based on DFT that $\textrm{Fe}_3\textrm{Sn}_2$ is a magnetic Weyl semimetal \cite{arXiv.1810.01514}. 

Two massive Dirac fermionic states around the $\bar{K}$ points have been previously observed by VUV-ARPES \cite{Ye2018}, and the same results are indeed reproduced by our VUV-ARPES measurements [Fig.\ \ref{Figure: Ek, DFT}(e)]. 
The unique structures have been assumed to originate from bilayer stacking with weak interlayer coupling.
However, we emphasize here that these features are not obtained in DFT band dispersions [Fig.\ \ref{Figure: Ek, DFT}(g)], no matter what $k_z$ value is selected [Fig.\ S9 \cite{Supplemental}].
At most, only one pair of cones is obtained in the DFT calculations, and such an electronic structure is also confirmed in our bulk sensitive SX-ARPES data [dashed curve in Fig.\ \ref{Figure: Ek, DFT}(f)].
While the steepness of the cones in Figs.\ \ref{Figure: Ek, DFT}(f) and \ref{Figure: Ek, DFT}(g) seems to be different from each other, this circumstance would not be critical for our arguments since the scaling and shift of band dispersions in the energy direction often become required to get a good agreement between DFT calculations and ARPES experiments \cite{Kuroda2017,Liu1282}.
Most importantly, the fact that two Dirac states with negligible $k_z$ dispersions are observed only by VUV-ARPES [Fig.\ \ref{Figure: Ek, DFT}(e)] leads us to conclude that these dispersions are derived dominantly from the surface states, which are unlikely related to bulk phenomena such as the AHE.

In conclusion, we combined two experimental techniques of surface-sensitive VUV-ARPES and relatively bulk-sensitive SX-ARPES, and separately observed two-dimensional states on the surface and three-dimensional states in the bulk of the kagome bilayer crystal $\textrm{Fe}_3\textrm{Sn}_2$. The crystal structure with the rhombohedral primitive unit cell allowed us to separate two- and three-dimensional states by observing the periodicity of ARPES intensities on a plane parallel to the $k_x$-$k_y$ plane as well as in the $k_z$ dispersion. We reveal that the massive Dirac fermionic states, which have been suggested to be the origin of the AHE, have dominant contribution from the surface states. In addition, we argue the requirement of taking into account the breathing effect on the degeneration property and the Berry curvature (Notes 1--3 \cite{Supplemental}) in the physics of the kagome lattice.
Moreover, the rhombohedral unit cell of $\textrm{Fe}_3\textrm{Sn}_2$ and the observed three-dimensional electronic structure make it not straightforward to discuss the properties of this material based on the two-dimensional kagome physics. One possible explanation for the large AHE  is that $\textrm{Fe}_3\textrm{Sn}_2$ is a magnetic Weyl semimetal. A good agreement between bulk states observed by SX-ARPES and band dispersions obtained by DFT calculations supports this perspective.

\begin{acknowledgments}
We thank Y. Ishida for supporting the analysis of our ARPES data \cite{doi:10.1063/1.5007226}.
This work is also supported by Grants-in-Aid for Scientific Research (B) (Grants No. 18H01165 and No. 19H02683) and Photon and Quantum Basic Research Coordinated Development Program from MEXT.
The SX synchrotron radiation experiments were performed with the approval of JASRI (Proposals No.\ 2019A1087 and No.\ 2019B1092).
We thank Diamond Light Source for access to beamline I05 (SI24488-1) that contributed to the results presented here.
\end{acknowledgments}


\bibliography{Fe3Sn2_ARPES_reference}

\end{document}



\title{Supplemental Material:\\Three-dimensional electronic structure in ferromagnetic $\textrm{Fe}_3\textrm{Sn}_2$ with breathing kagome bilayers}


\author{Hiroaki Tanaka}
\affiliation{ISSP, University of Tokyo, Kashiwa, Chiba 277-8581, Japan}

\author{Yuita Fujisawa}
\affiliation{Okinawa Institute of Science and Technology Graduate University, Okinawa 904-0495, Japan}

\author{Kenta Kuroda}
\email{kuroken224@issp.u-tokyo.ac.jp}
\affiliation{ISSP, University of Tokyo, Kashiwa, Chiba 277-8581, Japan}

\author{Ryo Noguchi}
\affiliation{ISSP, University of Tokyo, Kashiwa, Chiba 277-8581, Japan}

\author{Shunsuke Sakuragi}
\affiliation{ISSP, University of Tokyo, Kashiwa, Chiba 277-8581, Japan}

\author{C\'edric Bareille}
\affiliation{ISSP, University of Tokyo, Kashiwa, Chiba 277-8581, Japan}

\author{Barnaby Smith}
\affiliation{Okinawa Institute of Science and Technology Graduate University, Okinawa 904-0495, Japan}

\author{Cephise Cacho}
\affiliation{Diamond Light Source, Harwell Campus, Didcot OX11 0DE, United Kingdom}

\author{Sung Won Jung}
\affiliation{Diamond Light Source, Harwell Campus, Didcot OX11 0DE, United Kingdom}

\author{Takayuki Muro}
\affiliation{Japan Synchrotron Radiation Research Institute (JASRI), 1-1-1 Kouto, Sayo, Hyogo 679-5198, Japan}

\author{Yoshinori Okada}
\affiliation{Okinawa Institute of Science and Technology Graduate University, Okinawa 904-0495, Japan}

\author{Takeshi Kondo}
\email{kondo1215@issp.u-tokyo.ac.jp}
\affiliation{ISSP, University of Tokyo, Kashiwa, Chiba 277-8581, Japan}
\affiliation{Trans-scale Quantum Science Institute, University of Tokyo, Bunkyo-ku, Tokyo 113-0033, Japan}





\maketitle


\renewcommand{\thefigure}{S\arabic{figure}}
\renewcommand{\thetable}{S\Roman{table}}

\titleformat*{\section}{\bfseries}

\section*{Note 1. Tight-binding model and group theoretical analysis of the kagome lattice}
\begin{figure*}[bh]
\includegraphics{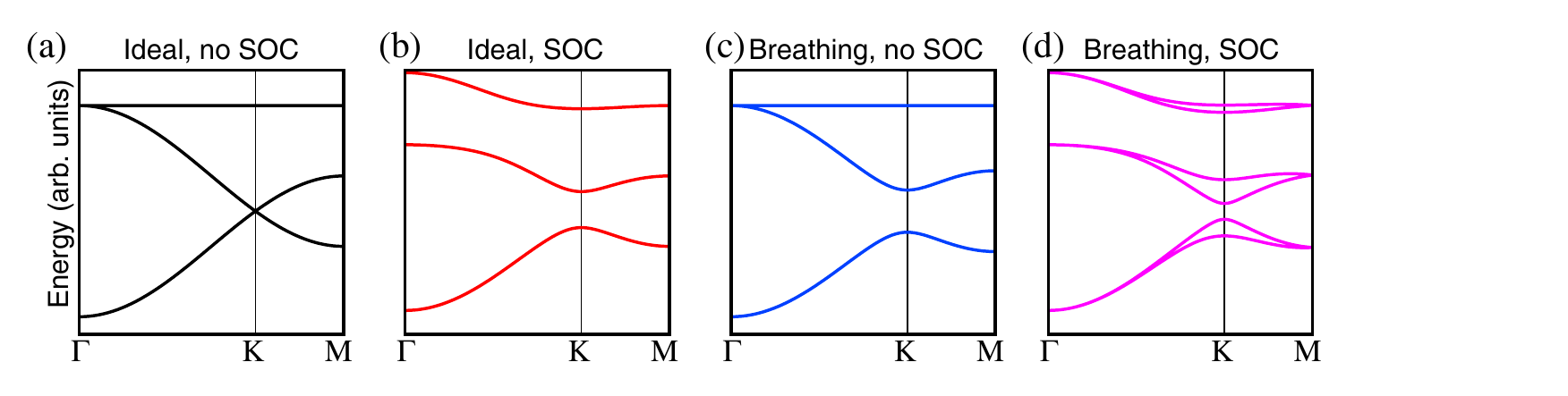}
\caption{\label{Figure: Tight-binding} Calculated band dispersions using the tight-binding model of (a) ideal kagome lattice without SOC, (b) ideal kagome lattice with SOC, (c) breathing kagome lattice without SOC, and (d) breathing kagome lattice with SOC. The breathing parameter $\beta$ is equal to 0.2 in (c), 0.05 in (d), and 0 otherwise. The SOC parameter $\phi$ is equal to $-0.3$ when SOC is on, and 0 otherwise. The SOC angle parameter $\alpha$ is $\pi/2$ (intrinsic SOC) in the whole cases.}
\end{figure*}

In this Note, we discuss the effects of the breathing and spin-orbit coupling (SOC) on the kagome lattice.
First, we calculate the band dispersions of the kagome lattice using the tight-binding model in Ref.\ [24].
After the Fourier transformation of eq.\ (4) in Ref.\ [24], we obtain the following $6\times6$ matrix Hamiltonian:
\begin{equation}
\mathcal{H}({\mathbf k})=-\begin{pmatrix}
0 & \hat{t}_1^\dagger e^{i\mathbf{k}\cdot\mathbf{a}_1/2}+\hat{t}_1^{\prime\dagger}e^{-i\mathbf{k}\cdot\mathbf{a}_1/2} & \hat{t}_3e^{-i\mathbf{k}\cdot\mathbf{a}_3/2}+\hat{t}_3^{\prime}e^{i\mathbf{k}\cdot\mathbf{a}_3/2}\\
\hat{t}_1e^{-i\mathbf{k}\cdot\mathbf{a}_1/2}+\hat{t}_1^\prime e^{i\mathbf{k}\cdot\mathbf{a}_1/2} & 0 & \hat{t}_2^\dagger e^{i\mathbf{k}\cdot\mathbf{a}_2/2}+\hat{t}_2^{\prime\dagger} e^{-i\mathbf{k}\cdot\mathbf{a}_2/2}\\
\hat{t}_3^\dagger e^{i\mathbf{k}\cdot\mathbf{a}_3}+\hat{t}_3^{\prime\dagger} e^{-i\mathbf{k}\cdot\mathbf{a}_3/2} & \hat{t}_2e^{-i\mathbf{k}\cdot\mathbf{a}_2/2}+\hat{t}_2^\prime e^{i\mathbf{k}\cdot\mathbf{a}_2/2} & 0
\end{pmatrix}
\label{Equation: kagome_Hamiltonian}
\end{equation}
where the vectors ${\mathbf a}_j$ and the hopping operators $\hat{t}_j$ and $\hat{t}_j^\prime$ are defined in the reference $(j=1,\ 2,\ 3)$.
We introduce new parameters $\beta=\delta/t_0$ and $t_\mathrm{total}=\sqrt{t_0^2+\lambda_{0,\textrm{i}}^2+\lambda_{0,\textrm{R}}^2}$ to simplify the equations, then the hopping operators $\hat{t}_j$ and $\hat{t}_j^\prime$ can be represented as follows:
\[\hat{t}_j=(1+\beta)t_\mathrm{total}\exp(i\phi\hat{\mathbf{d}}_j\cdot{\boldsymbol \sigma}),\ \hat{t}_j^\prime=(1-\beta)t_\mathrm{total}\exp(i\phi\hat{\mathbf{d}}_j\cdot{\boldsymbol \sigma})\]
$\beta$ and $\phi$ correspond to the strengths of the breathing and SOC, respectively.
The band dispersions of several types of kagome lattices are summarized in Fig.\ \ref{Figure: Tight-binding}.

Tables \ref{Table: Irreducible_ideal} and \ref{Table: Irreducible_breathing} show how the degenerations appear and disappear at highly symmetric $\Gamma$, K, and M points in the kagome lattice.
In the following argument based on Refs. [17, 26-28], we consider two-dimensional point groups.
In two dimensions, the ideal kagome lattice has 6-fold rotational symmetry and six mirror planes ($6mm$) and the breathing lattice has 3-fold rotational symmetry and three mirror planes ($3m$).
Figures \ref{Figure: Tight-binding}(a) and (c) correspond to the spin neglected cases (three eigenstates) and \ref{Figure: Tight-binding}(b) and (d) correspond to the spin included cases (six eigenstates).
We note that in Fig.\ \ref{Figure: Tight-binding}(b) double degenerations occur anywhere nonetheless with the inclusion of SOC.

\begin{table}
	\caption{Irreducible representations of the ideal kagome lattice}
	\label{Table: Irreducible_ideal}
	\begin{tabular}{ccc}\hline\hline
	Point& Spin Neglected & Spin Included \\\hline
	$\Gamma$ $(6mm)$ & $A_1+E_2$ & $E_{1/2}+E_{3/2}+E_{5/2}$\\
	K $(3m)$ & $A_1+E$ & $2E_{1/2}+E_{3/2}$ \\
	M $(2mm)$ & $A_1+B_1+B_2$ & $3E_{1/2}$\\\hline\hline
	\end{tabular}
\end{table}

\begin{table}
	\caption{Irreducible representations of the breathing kagome lattice}
	\label{Table: Irreducible_breathing}
	\begin{tabular}{ccc}\hline\hline
	Point & Spin Neglected & Spin Included \\\hline
	$\Gamma\ (3m)$ & $A_1+E$ & $2E_{1/2}+E_{3/2}$\\
	K $(3)$ & $A+E$ & $2E_{1/2}+2B_{3/2}$ \\
	M $(m)$ & $2A^\prime+A^{\prime\prime}$ & $3E_{1/2}$\\\hline\hline
	\end{tabular}
\end{table}

We use, as basis functions, wave functions centered at each atom in the kagome unit cell.
In the first columns of the tables, the parentheses on the right of each point label enclose the symbol of the point group of the operations which keep the point invariant in the reciprocal space.
The characters of the representation of our basis functions are calculated and then the representation is reduced to several irreducible representations as shown in Tables \ref{Table: Irreducible_ideal} and \ref{Table: Irreducible_breathing}.

An irreducible representation $E$ (sometimes with subscripts) is two-dimensional, which generally means a double degeneration.
The exception of the degeneration is the case where $E$ can be reduced to two one-dimensional irreducible representations in the range of complex numbers and the Kramers degeneration does not occur.
Since external field is not applied, time-reversal symmetry (TRS) allows the Kramers degeneration.
However, for the additional degeneration originating from the TRS to occur in band dispersions, the lattice needs to have the space inversion symmetry or the high-symmetric point needs to be one of the time-reversal invariant momenta (TRIM).
Since the ideal kagome lattice has the space inversion symmetry and the $\Gamma$ and M points are TRIM, physically irreducible two-dimensional representations cause degenerations in these situations.
The K points in the breathing kagome lattice do not satisfy the above two conditions, and therefore the degeneration does not occur.
With these considerations, the appearance and disappearance of the degenerations represented in Fig. \ref{Figure: Tight-binding} can be explained.

\section*{Note 2. Berry curvature of the TRS-breaking kagome lattice}
\begin{figure*}[hb]
\includegraphics{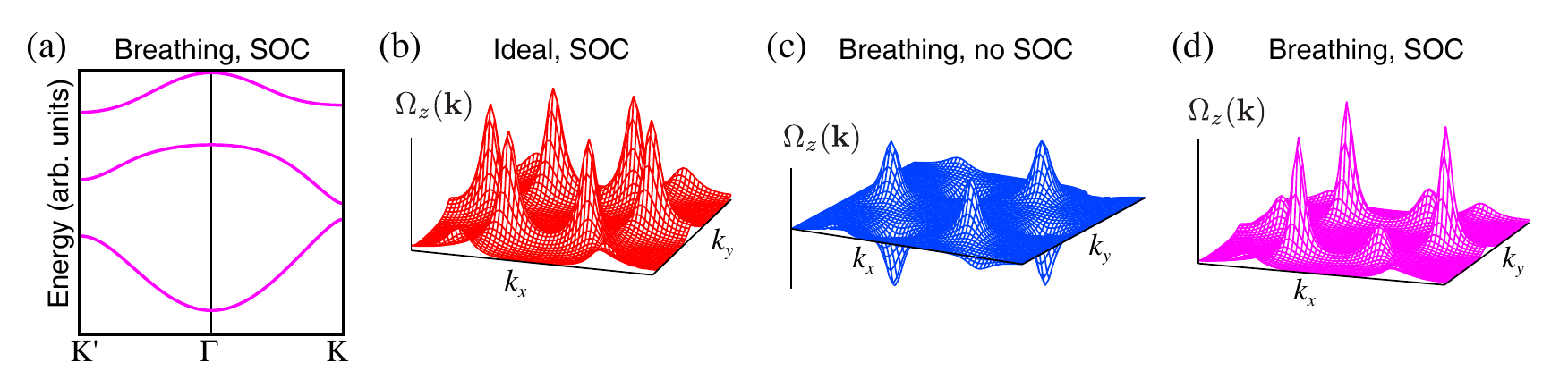}
\caption{\label{Figure: Berry-curvature} Berry curvature of the kagome lattice. (a) Band dispersions of the $3\times3$ Hamiltonian with the inclusion of the breathing and SOC. (b) Berry curvature (in arbitrary units) of the ideal kagome lattice with SOC, (c) that of the breathing kagome lattice without SOC, and (d) that of the breathing kagome lattice with SOC. Local maxima or minima are at the K or $\textrm{K}^\prime$ points. $\beta$ is set to 0.2 in (c) and 0.05 in (d) and $\phi$ is set to $-0.3$ in (b) and (d).}
\end{figure*}

To calculate the Berry curvature, we now break the TRS by some external magnetic field.
As a result, only one direction of spin (up spin is used here) is considered.
We compose the $3\times3$ Hamiltonian from eq.\ (\ref{Equation: kagome_Hamiltonian}) by extracting the odd lines and the odd columns.
In the cases where either one of SOC and the breathing is included, the band dispersions of the $3\times3$ Hamiltonian is the same as Figs. \ref{Figure: Tight-binding}(b) or (c).
When both are included, the gap sizes at the K and $\textrm{K}^\prime$ points become different [Fig.\ \ref{Figure: Berry-curvature}(a)], where the $\textrm{K}^\prime$ point is the symmetric position of the K point with respect to the $\Gamma$ point.

The Berry curvature of the lowest band is calculated according to the definition $\mathbf{\Omega}(\mathbf{k})=i(\nabla_{\mathbf{k}}\psi^*(\mathbf{k}))\times(\nabla_{\mathbf{k}}\psi(\mathbf{k}))$ [2], where $\psi(\mathbf{k})$ is the eigenstate of the lowest band.
Figures \ref{Figure: Berry-curvature}(b), (c), and (d) show the $z$ components of the Berry curvatures.
The Berry curvature caused by SOC is an even function with Chern number 1, while that caused by the breathing is odd with Chern number 0.
When both are included, the difference of the gap sizes appears in the magnitude of the Berry curvature.

\section*{Note 3. Group theoretical analysis of the breathing kagome bilayer}
We obtain degeneration properties of the breathing kagome bilayer based on group theory.
In this Note, we neglect the spin and SOC and calculate the representations with six basis functions of the bilayer lattice, similarly to the previous research about ferromagnetic kagome bilayer [20].

\begin{figure*}[hb]
\includegraphics{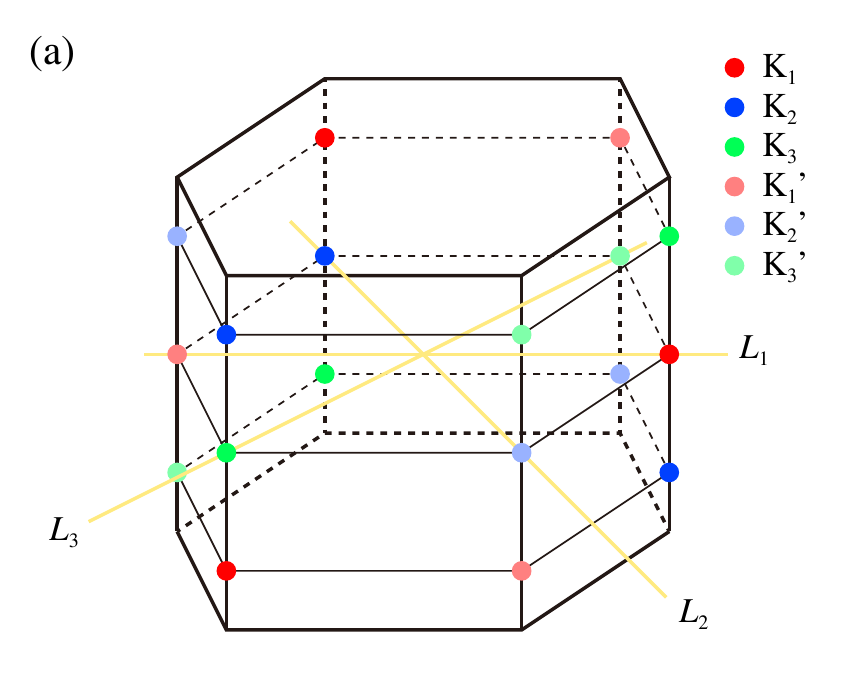}
\caption{\label{Figure: Kpoints} (a) Distribution of the K points. 2-fold rotational operation along the axis $L_i$ keeps the $\textrm{K}_i$ and $\textrm{K}_i^\prime$ invariant $(i=1,\ 2,\ 3)$.}
\end{figure*}

First, we use the primitive rhombohedral unit cell in the real space and the hexagonal column unit cell in the reciprocal space, which corresponds to the periodicity of ARPES spectra.
Because of the unusual stacking of the hexagonal column unit cells discussed in the main text, all K points are treated as different (but equivalent) points [Fig.\ \ref{Figure: Kpoints}].
For example, $\left(\dfrac{4\pi}{3a},0,0\right)$, $\left(-\dfrac{2\pi}{3a},\dfrac{2\pi}{\sqrt{3}a},\dfrac{2\pi}{c}\right)$, and $\left(-\dfrac{2\pi}{3a},-\dfrac{2\pi}{\sqrt{3}a},-\dfrac{2\pi}{c}\right)$ are the same points labeled by $\textrm{K}_1$.
Therefore, the K points are not invariant under 3-fold rotations.
In the space group $R\bar{3}m$, only a 2-fold rotational operation along an axis on the $k_xk_y$ plane keeps the K points invariant [Fig.\ \ref{Figure: Kpoints}], then the representation is reduced to $3A+3B$ (no degeneration).

\begin{table}
	\caption{Irreducible representations of the breathing kagome monolayer and bilayer. Parentheses in the first column encloses the symbols of the space groups of the bilayer.}
	\label{Table: Irreducible_breathing_conventional}
	\begin{tabular}{ccc}\hline\hline
	Point & Monolayer & Bilayer\\\hline
	$\Gamma\ (\bar{3}m)$ & $A_1+E$ & $(A_{1g}+A_{2u})+(E_g+E_u)$\\
	K $(32)$ & $A+E$ & $(A_1+A_2)+2E$ \\
	M $(2/m)$ & $2A^\prime+A^{\prime\prime}$ & $2(A_g+B_u)+(B_g+A_u)$ \\\hline\hline
	\end{tabular}
\end{table}

\begin{figure*}

\includegraphics{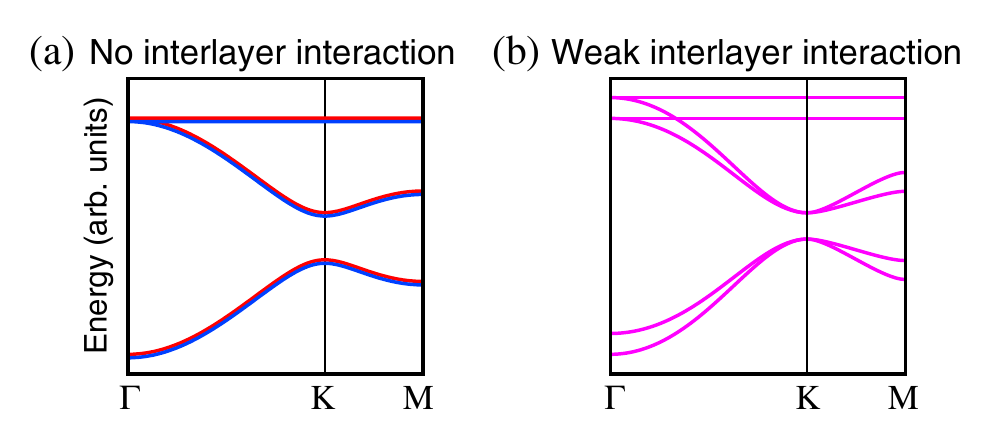}
\caption{\label{Figure: Band_breathing_conventional} Schematics of band dispersions of the breathing kagome bilayer when the conventional unit cell is used. (a) Schematic of band dispersions of breathing kagome bilayer with no interlayer interaction and (b) that with weak interaction, keeping degeneration properties obtained from group theoretical analysis. The band crossing between the $\Gamma$ and K points is not systematic and thus may not exist.}
\end{figure*}

If we use the conventional unit cell, the reciprocal unit cell is folded to a hexagonal column with a height of $2\pi/c$.
The three-dimensional space group of a high-symmetry point includes the two-dimensional space group, so we can extend two-dimensional irreducible representations [Table\ \ref{Table: Irreducible_breathing}] to three-dimensional ones.
Table \ref{Table: Irreducible_breathing_conventional} shows degeneration properties at the $\Gamma$, K, and M points.
Actually two degenerations occur at the K points, but if we assume weak interlayer interaction, the degenerated bands will be parabolic [Fig. \ref{Figure: Band_breathing_conventional}(b)].
Therefore, these degenerations do not necessarily mean the vanishment of the breathing effect.

\clearpage

\section*{Note 4. Sample growth and X-ray diffraction}
\begin{figure}[h]
\includegraphics{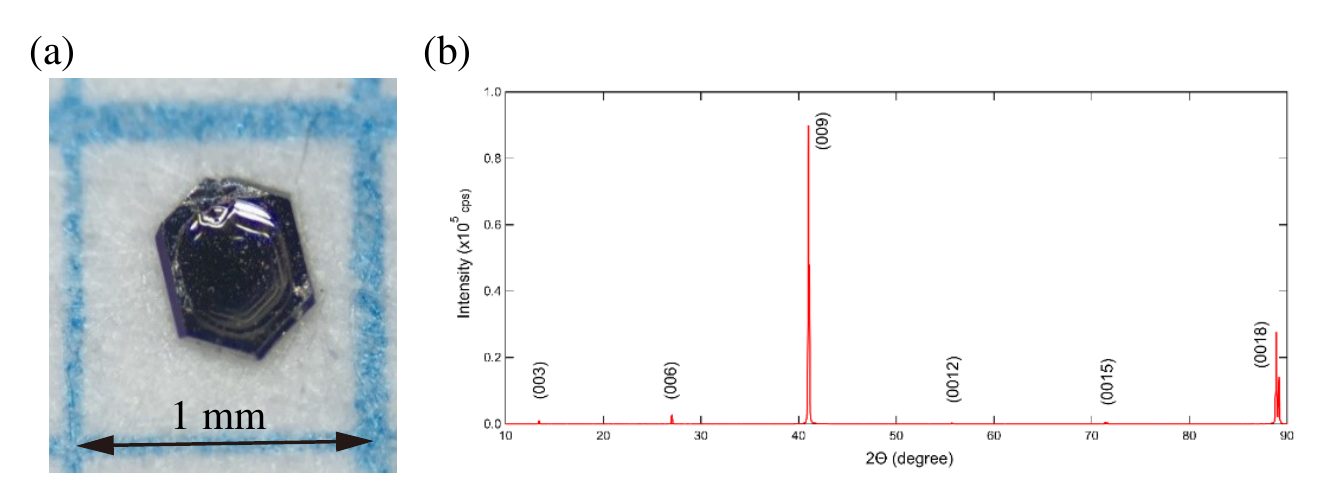}
\caption{\label{Figure: Growth} Growth and characterization of $\textrm{Fe}_3\textrm{Sn}_2$ single crystals. (a) Picture of a typical crystal with a clear hexagonal shape. (b) Out of plane X-ray diffraction pattern with clear $(00l)$ peaks.}
\end{figure}

The samples were grown by chemical vapor transport. 
Figure \ref{Figure: Growth}(a) shows a typical image of an obtained crystal, showing clear plate-like morphology with a hexagonal shape.

The crystals were characterized by X-ray diffraction (XRD) using a Bruker D8 Venture (single crystal) and a Bruker D8 Discover (out of plane).
The lattice constants $a=b=5.35\ \textrm{\AA}$ and $c=19.82\ \textrm{\AA}$ were determined based on single crystal XRD data.
Figure \ref{Figure: Growth}(b) shows the out of plane diffraction $(00l)$.
There are no secondary crystal phases and all peaks are indexed in accordance with the expected $\textrm{Fe}_3\textrm{Sn}_2$ crystal.

\section*{Note 5. Bulk states observed by VUV-ARPES}
\begin{figure*}[hb]
\includegraphics{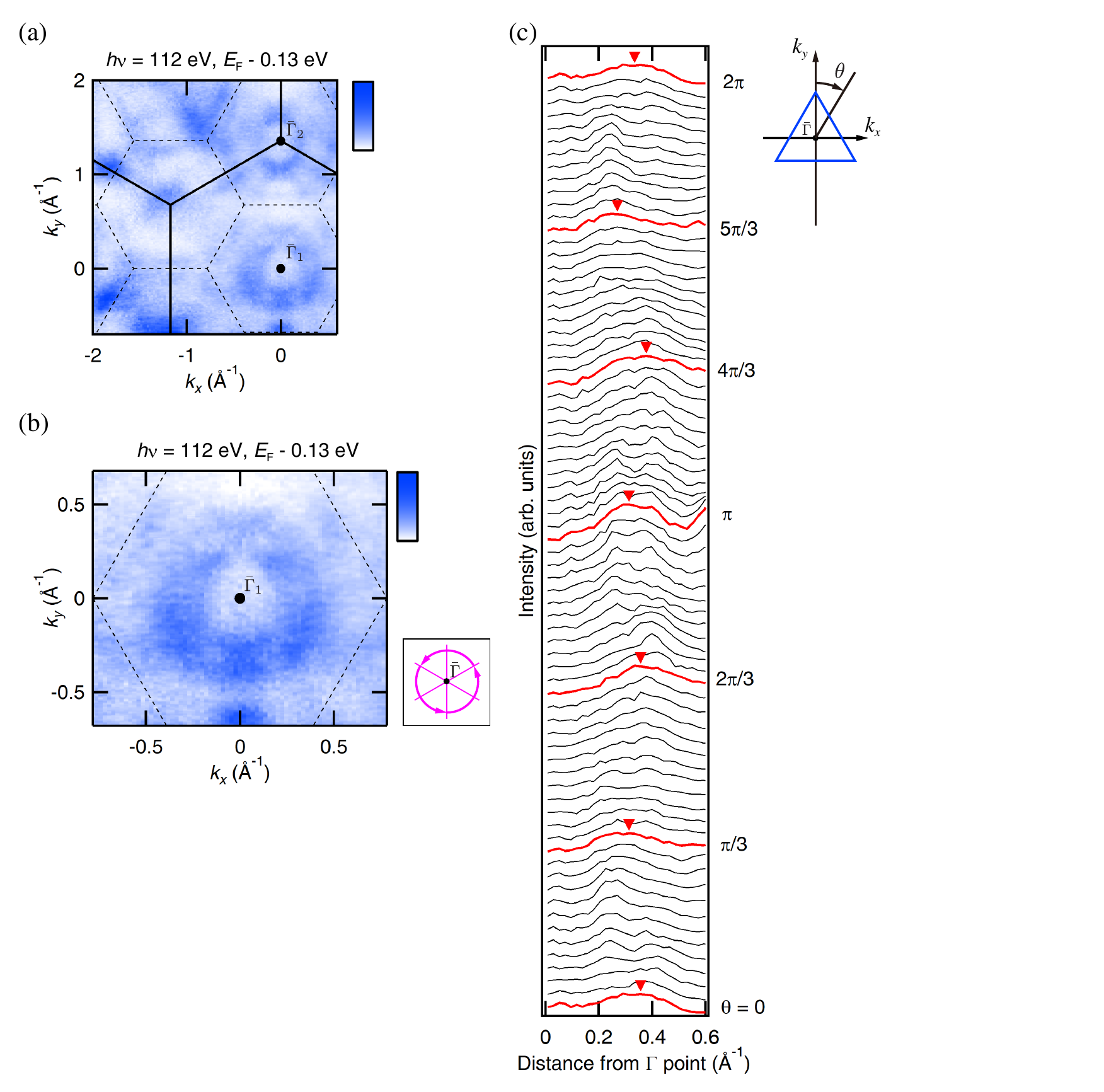}
\caption{\label{Figure: Bulk-VUV} Bulk states measured by VUV light. (a) Constant-energy mapping measured by 112 eV VUV light. The solid lines are the periodicity of the three-dimensional states and the dashed lines are that of the two-dimensional states. (b) Triangular-shaped pocket centered at the $\bar{\Gamma}_1$ point. The pink arrow arcs and lines in the bottom right corner represent 3-fold rotational symmetry and three mirror planes of the bulk states. (c) MDCs with different rotational angles $\theta$. The schematic in the top right corner shows the definition of $\theta$.}
\end{figure*}

Figure \ref{Figure: Bulk-VUV} shows bulk bands observed by VUV-ARPES.
In the constant-energy mapping [Fig.\ \ref{Figure: Bulk-VUV}(a)],  dispersions show a long periodicity of the three-dimensional states.
Additionally, the triangular-shaped pocket centered at the $\bar{\Gamma}_1$  point [Fig.\ \ref{Figure: Bulk-VUV}(b)] reflects 3-fold rotational symmetry and mirror planes originating from the three-dimensionality.
The graph of the MDCs with different rotational angles [Fig.\ \ref{Figure: Bulk-VUV}(c)] shows three cycles of the change of the peak positions  in $2\pi$ rotation, which supports the triangular shape of the pocket.

\clearpage
\section*{Note 6. Negligible $k_z$ dispersion measured by VUV-ARPES}
\begin{figure*}[hb]
\includegraphics{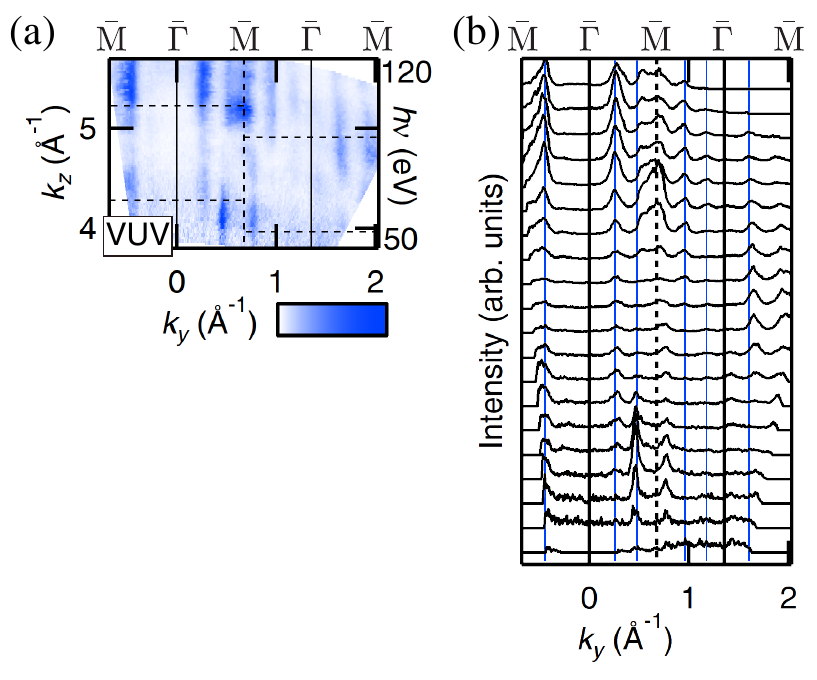}
\caption{\label{Figure: kz-VUV} Negligible $k_z$ dispersion measured by VUV-ARPES. (a) $k_y$-$k_z$ mapping at the Fermi energy measured by VUV light [same as Fig.\ 3(a)]. (b) Momentum distribution curves of (a), with overlaid vertical blue lines.}
\end{figure*}

Figure \ref{Figure: kz-VUV}(b) shows momentum distribution curves corresponding to the $k_y$-$k_z$ mapping taken by VUV-ARPES [Fig.\ \ref{Figure: kz-VUV}(a)].
Although the matrix element effects weakened the intensity of several peaks, one can see negligible $k_z$ dispersion along overlaid blue lines.
\clearpage

\section*{Note 7. Curvature plot of SX-ARPES mapping}
\begin{figure*}[hb]
\includegraphics{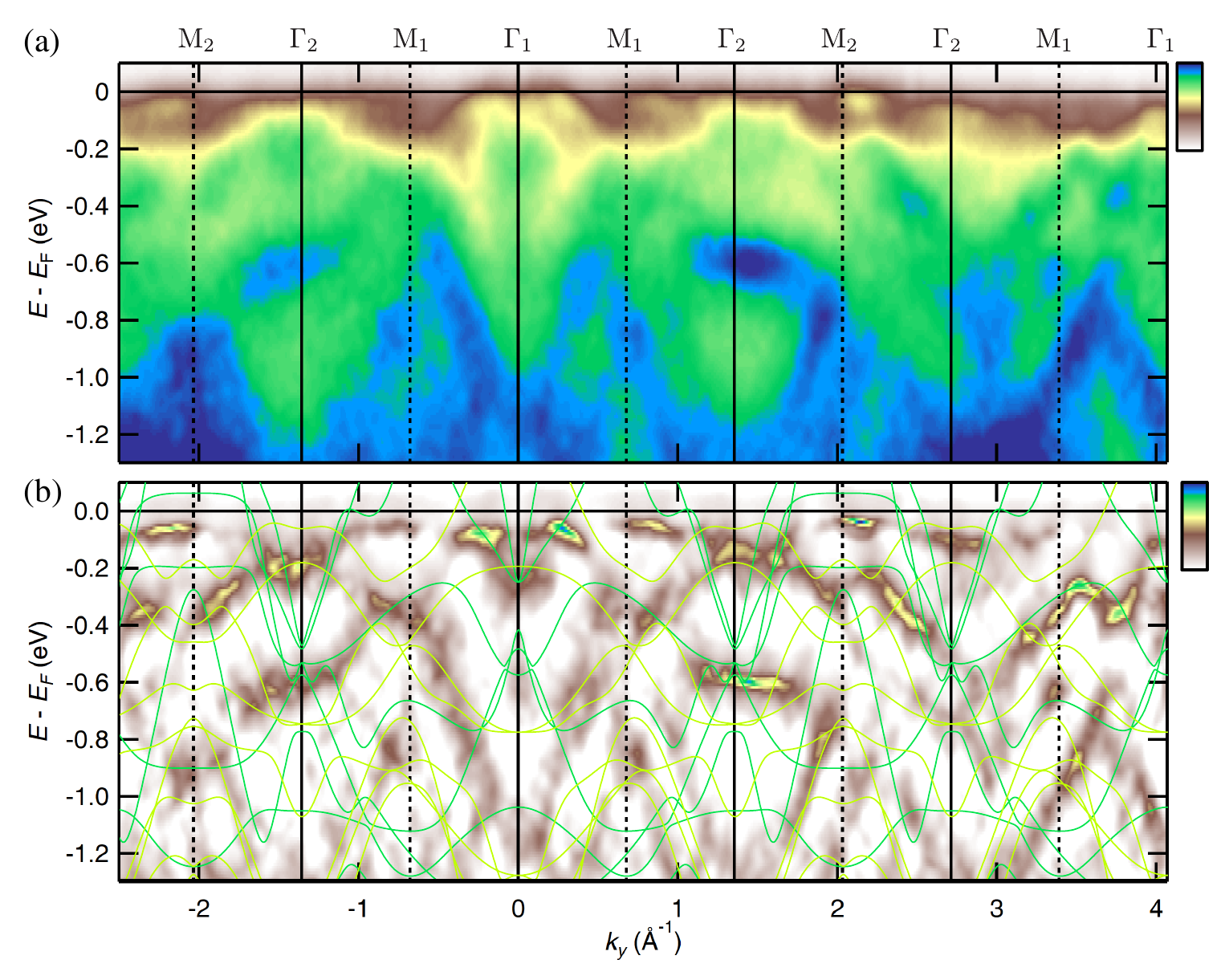}
\caption{\label{Figure: Curvature} Enhanced band dispersions by curvature plot. (a) Original SX-ARPES mapping along the $k_y$ direction [same as Fig.\ 4(a)]. (b) Curvature plot of (a), with overlaid DFT band dispersions.}
\end{figure*}

We used curvature plot [38], which can enhance several band dispersions, to the SX-ARPES mapping shown in Fig.\ \ref{Figure: Curvature}(a).
Figure \ref{Figure: Curvature}(b) shows the enhanced bulk band dispersions.

\clearpage
\section*{Note 8. DFT band dispersions near $\bar{\textrm{K}}$ point}
\begin{figure*}[hb]
\includegraphics{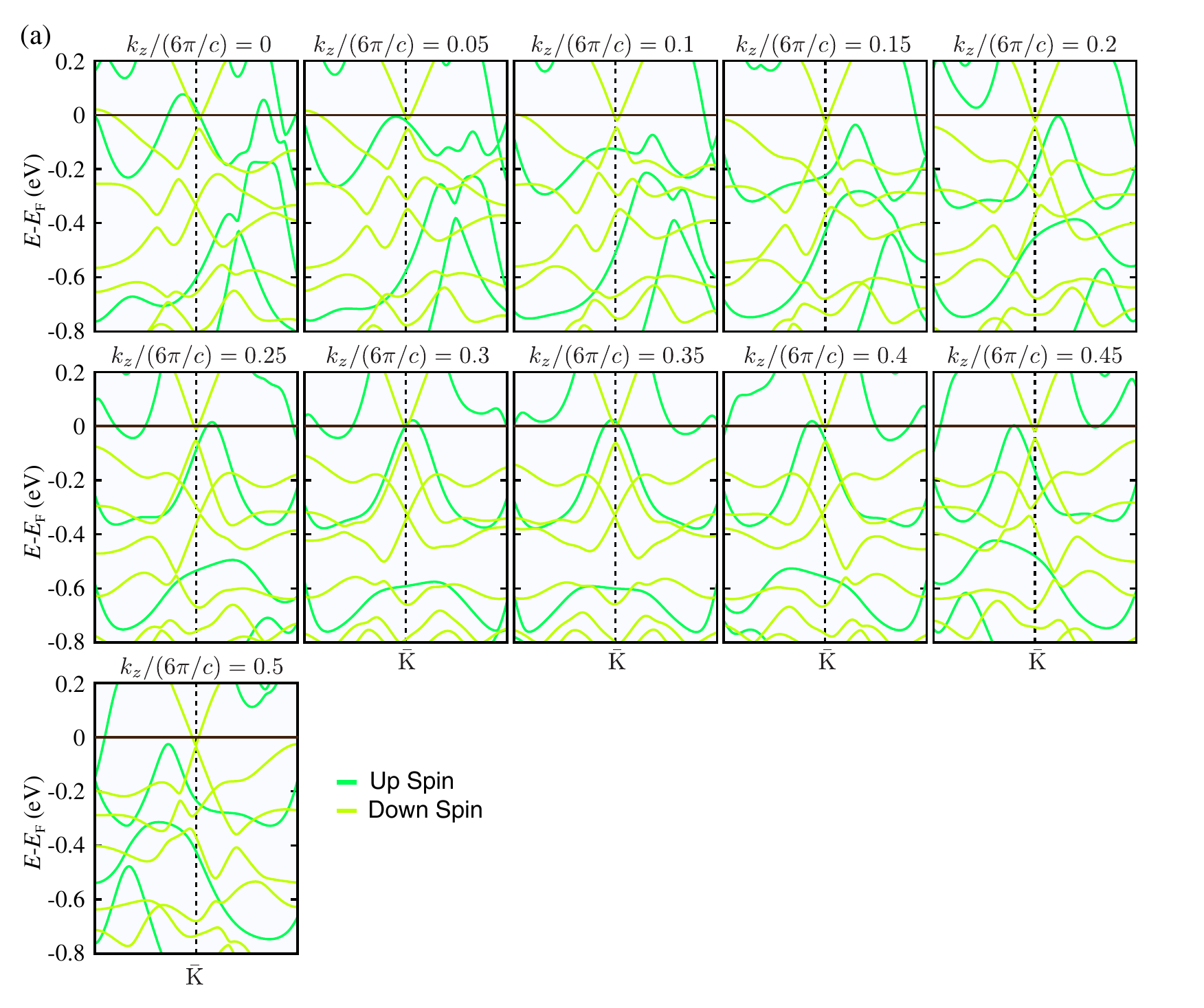}
\caption{\label{Figure: DFT-KKK} DFT band dispersions near the $\bar{\textrm{K}}$ point with the difference of the $k_z$ coordinate.}
\end{figure*}

Figure \ref{Figure: DFT-KKK} shows DFT band dispersions near the $\bar{\textrm{K}}$ point from $k_z=0$ (the $\Gamma$ point is on the plane) to $k_z=3\pi/c$ (the Z point is on the plane). The horizontal axes are the same as that of Fig.\ 4(g) in the main text.

